\documentclass[smallextended]{svjour3}       % onecolumn (second %format)

\smartqed  % flush right qed marks, e.g. at end of proof
\usepackage{amsfonts}
\usepackage{amsmath}
\usepackage{amssymb}
\usepackage{graphicx}
\usepackage{latexsym}
\usepackage{color}%
\begin{document}

\title{New scenarios for classical and quantum mechanical systems with position dependent mass}

%\subtitle{New scenarios ...}

\titlerunning{New scenarios ...}        % if too long for running head

\author{J. R. Morris}


\institute{Department of Physics, Indiana University Northwest, 3400 Broadway, Gary, IN
46408, USA\\email: jmorris@iun.edu}

%The correct dates will be entered by the editor

\maketitle

\begin{abstract}
An inhomogeneous Kaluza-Klein compactification to four dimensions, followed by
a conformal transformation, results in a system with position dependent mass
(PDM). This origin of a PDM is quite different from the condensed matter one.
A substantial generalization of a previously studied nonlinear oscillator with
variable mass is obtained, wherein the position dependence of the mass of a
nonrelativistic particle is due to a dilatonic coupling function emerging from
the extra dimension. Previously obtained solutions for such systems can be
extended and reinterpreted as nonrelativistic particles interacting with
dilaton fields, which, themselves, can have interesting structures. An
application is presented for the nonlinear oscillator, where within the new
scenario the particle is coupled to a dilatonic string.

\PACS{03.65.Ca \and 11.10.Kk \and 04.50.Cd \and 11.27.+d}

%\pacs{03.65.Ca, 11.10.Kk, 04.50.Cd, 11.27.+d}
\keywords{position dependent mass \and nonlinear quantum oscillator \and Kaluza-Klein theory \and dimensional reduction}
\maketitle

\subclass{70Gxx \and 70Hxx \and 70Sxx \and 81Sxx}
\end{abstract}

\section{Introduction}

\ \ \ Quantum mechanical systems involving nonrelativistic particles with
position dependent masses have been studied by numerous authors (see, e.g.,
\cite{Mathews}-\cite{epjp}). Such systems can arise in condensed matter
settings, as pointed out, for example, in Ref.\cite{cari}, but are of
mathematical interests in their own right. An example of such a system is a
previously studied prototype\cite{Mathews},\cite{cari} that derives from a
classical lagrangian%
\begin{equation}
L=\frac{1}{2}\frac{m_{0}}{(\lambda x^{2}+1)}\left(  \dot{x}^{2}-\alpha
^{2}x^{2}\right)  \label{a1}%
\end{equation}

which can also be extended to higher dimensions\cite{refa},\cite{refb}%
,\cite{refd},\cite{jmp},\cite{epjp}. The particle described by (\ref{a1}) has
an effective position dependent mass (PDM)%
\begin{equation}
m(x)=\frac{m_{0}}{(\lambda x^{2}+1)}\label{a2}%
\end{equation}

and is subject to an effective \textquotedblleft spring
constant\textquotedblright\ $k(x)=m(x)\alpha^{2}$. This problem has been
studied at both the classical and quantum mechanical level, and in various
space dimensions. (See, e.g., \cite{cari} and \cite{axel1}-\cite{epjp} for
exact solutions of the quantum mechanical problem.)

\bigskip

\ \ One problem that arises for systems with a PDM is the question of how to
proceed from a classical description to a quantum one, since an ordering
ambiguity for momentum operators becomes apparent. The quantization can be
accomplished with several approaches\cite{young},\cite{cari},\cite{refe}. One
approach, that has been used in several studies involving quantum mechanical
particles with position dependent masses, is one where the classical kinetic
energy is factorized into two symmetric parts, each involving a canonical
momentum. The classical momentum is then canonically replaced with its quantum
counterpart, $\vec{p}\rightarrow-i\hbar\nabla$. A classical kinetic energy of
the form $T=\frac{1}{2m_{0}}f^{-1}(\vec{x})\vec{p}\cdot\vec{p}=\frac{1}%
{2m_{0}}(f^{-1/2}\vec{p})\cdot(f^{-1/2}\vec{p})$ then takes its quantum
form\cite{cari},\cite{axel1},\cite{epjp}%
\begin{equation}
T\rightarrow\hat{T}=-\frac{\hbar^{2}}{2m_{0}}[f^{-1/2}(\vec{x})\nabla
]\cdot\lbrack f^{-1/2}(\vec{x})\nabla]\label{0}%
\end{equation}

where $m_{0}$ is a constant mass related to an effective position dependent
mass $m(\vec{x})$ by $m(\vec{x})=f(\vec{x})m_{0}$. Other approaches exist as
well, and different quantization prescriptions can result in different
Hamiltonians\cite{refe},\cite{young}.

\bigskip

\ \ \ \ We can generalize the system in (\ref{a1}) by writing a lagrangian in
the form (in cartesian coordinates)%
\begin{equation}
L=\frac{1}{2}m(x)\left[  \delta_{ij}u^{i}u^{j}-U_{0}(\vec{x})\right]
\label{3}%
\end{equation}

where $m(x)U_{0}(\vec{x})$ represents a nonlinear potential energy, in
general, and the summation convention is used. It is to be demonstrated here,
that this form of lagrangian emerges naturally from an inhomogeneous
Kaluza-Klein compactification of an extra spacetime dimension to yield an
effective 4D theory. The extra space dimension, when compactified, gives rise
to an effective mass $m(x)$ in the effective $4$-dimensional theory. An
inhomogeneous compactification is manifest as a spacetime dependent scale
factor $b(x^{\mu})$ for the extra dimension, so that there can be regions of
spacetime where the size of the extra dimension becomes larger or smaller.
This can result in some interesting physical effects and objects, for example,
gravitational bags\cite{bags}, dimension bubbles\cite{bubbles}, and scattering
from dimensional boundaries\cite{scatter}. However, interest here is focused
upon the emergence and treatment of a low energy quantum mechanical system
where a position dependent mass is involved. In the 4D theory the function
$b(x^{\mu})$ is related to a dilaton coupling function which determines how
strongly the dilaton field \ couples to matter particles.

\bigskip

\ \ We begin with an action describing a classical particle moving in a five
dimensional spacetime with pure Einstein gravity and the inclusion of a
possible cosmological constant $\Lambda$. We then dimensionally reduce this 5D
theory to an effective 4D one, where the extra dimensional scale factor $b(x)$
is related to a scalar field $\phi(x)$ that is nonminimally coupled to the 4D
Ricci scalar $\tilde{R}[\tilde{g}_{\mu\nu}]$. A conformal transformation from
the 4D Jordan frame with\ metric $\tilde{g}_{\mu\nu}$ to an Einstein frame
with\ metric $g_{\mu\nu}$ results in a representation where the action
contains an ordinary 4D Einstein-Hilbert term, along with a scalar field
$\phi$ derived from the scale factor $b$, and a classical matter action. In
the Einstein frame representation of the theory, the scale factor $b(x)$ is no
longer coupled to the Ricci scalar, but now becomes coupled to the matter
sector of the theory\cite{Dicke62},\cite{FMbook}.

\bigskip

\ \ Passing to the flat 4-dimensional spacetime limit with nonrelativistic
matter, the 4D action can be recast in the form of a low energy theory for
nonrelativistic matter, along with a relativistic scalar field with a
potential $V\sim\Lambda b^{-1}$. The equations of motion for the system are
obtained, and explicit, exact solutions for the scalar field are obtainable in
certain cases. The classical matter particle has an attendant variable mass
$m(\vec{x})$ in a flat spacetime.

\bigskip

\ \ A quantization of the Hamiltonian then yields a Schr\"{o}dinger-like
equation for the system. We use the quantization method described above, where
the kinetic energy operator is given by (\ref{0}),%
\begin{equation}
\hat{H}\psi(\vec{x})=\left[  \hat{T}+\mathcal{U}_{0}(\vec{x})\right]
\psi(\vec{x})=E\psi(\vec{x});\ \ \ \ \ \mathcal{U}_{0}(\vec{x})=\frac{1}%
{2}m(\vec{x})U_{0}(\vec{x})\label{4}%
\end{equation}

which is obtained from a quantization procedure for a classical
nonrelativistic particle characterized by a Lagrangian in a flat space
representation of the form given by (\ref{3}). An application of this method
using a particular solution of the dilaton field equation is seen to produce
the system of (\ref{a1}) and (\ref{a2}), generalized to two or three spatial
dimensions, with $x\rightarrow r=\sqrt{x^{2}+y^{2}}$ being a radial
coordinate. In addition, there is a nontrivial dilatonic field configuration
which couples to the particle oscillator. This problem has been examined
previously, but without any connection of the position dependent mass to a
dilaton field. In this example, the dilaton field has a structure resembling a
global cosmic string. Another difference is that the mass function $m(\vec
{x})$ is not an arbitrary input function, but is determined by the field
equation for the dilaton scalar $\phi$.

\section{The model}

\subsection{The 5D action and dimensional reduction}

\ \ We begin by assuming a 5D dimensional spacetime equipped with a metric%
\begin{equation}
d\tilde{s}_{D}^{2}=\tilde{g}_{MN}(x^{\mu},y)dx^{M}dx^{N}=\tilde{g}_{\mu\nu
}(x^{\alpha})dx^{\mu}dx^{\nu}-b^{2}(x^{\alpha})dy^{2}\label{5}%
\end{equation}

where the 4-dimensional metric $\tilde{g}_{\mu\nu}(x)$ has a negative
signature $(+,-,-,-)$. We use $\mu,\nu=0,1,2,3$ for the ordinary spacetime
indices and $M,N=0,1,2,3,5$. Absolute values of metric determinants are
denoted by $\tilde{g}_{5}=|\det\tilde{g}_{MN}|$, and $\tilde{g}=|\det\tilde
{g}_{\mu\nu}|$, so that $\tilde{g}_{5}=\tilde{g}b^{2}$. The action for the 5D
theory that includes gravitation and matter is%
\begin{equation}
S=\int d^{5}x\sqrt{\tilde{g}_{5}}\left\{  \frac{1}{2\kappa_{5}^{2}}\left[
\tilde{R}_{5}[\tilde{g}_{MN}]-2\Lambda\right]  \right\}  +S_{m}[\tilde{g}%
_{MN},\cdot\cdot\cdot]\label{6}%
\end{equation}

where $\Lambda$ is a cosmological constant, and $S_{m}$ is the matter action
which depends upon particle velocities and metric $\tilde{g}_{MN}$. For
instance, the matter action for a free classical particle is given by%
\begin{equation}
S_{m}=-\int m_{0}d\tilde{s}_{5}=-\int m_{0}\sqrt{\tilde{g}_{MN}\tilde{u}%
^{M}\tilde{u}^{N}}d\tilde{s}_{5}\label{7}%
\end{equation}

where $m_{0}$ is a constant mass parameter in the $D$-dimensional theory, and
$\tilde{u}^{M}=dx^{M}/d\tilde{s}_{5}$, and we have $\sqrt{\tilde{g}_{MN}%
\tilde{u}^{M}\tilde{u}^{N}}=1$ when evaluated on the particle worldline\ where
$d\tilde{s}_{5}^{2}=\tilde{g}_{MN}dx^{M}dx^{N}$ holds. We have a 5D
gravitational constant denoted by $G_{5}$ with $\kappa_{5}^{2}=8\pi G_{5}$. It
is related to the 4D gravitational constant $G$ by $G_{D}=V_{y}G$, and hence
$\kappa_{D}^{2}=V_{y}\kappa^{2}$, where $V_{y}$ is the \textquotedblleft
coordinate volume\textquotedblright\ of internal space, $V_{y}=\int dy=2\pi
R_{0}$. We assume that particle trajectories and fields are $y$ independent,
i.e., there are no Kaluza-Klein (KK) modes. Specifically, we assume that the
classical particle's trajectory is confined to the 4D spacetime, with $dy=0$
along the particle's path. Therefore, along the particle's trajectory we have
$d\tilde{s}_{5}^{2}|_{\text{path}}=d\tilde{s}^{2}=\tilde{g}_{\mu\nu}%
(x)dx^{\mu}dx^{\nu}$ with $\tilde{u}^{5}=0$.

\bigskip

\ \ We dimensionally reduce the 5D theory to an effective 4D theory by
performing an integration over the internal space with $d^{5}x=d^{4}xdy$ and
using $\sqrt{\tilde{g}_{5}}=\sqrt{\tilde{g}}b(x)$. The 5D Ricci scalar
$\tilde{R}[\tilde{g}_{MN}]=\tilde{g}^{MN}\tilde{R}_{MN}[\tilde{g}_{MN}]$, with
$\tilde{R}_{MN}$ the Ricci tensor, can be broken into a 4D Ricci scalar
$R[\tilde{g}_{\mu\nu}]$ plus terms involving the scale factor $b(x)$ and its
derivatives. The result is (see, e.g.,\cite{bubbles},\cite{CGHW}%
,\cite{MorrisGRG})%
\begin{equation}%
\begin{array}
[c]{ll}%
S= & \int d^{4}x\sqrt{\tilde{g}}\{\dfrac{1}{2\kappa^{2}}\Big[b\tilde{R}%
[\tilde{g}_{\mu\nu}]+2\tilde{g}^{\mu\nu}\tilde{\nabla}_{\mu}\tilde{\nabla
}_{\nu}b-b\Lambda\Big]\}+S_{m}%
\end{array}
\label{8}%
\end{equation}

where $\tilde{\nabla}_{\mu}$ represents a covariant derivative with respect to
the metric $\tilde{g}_{\mu\nu}$. The matter action for a particle of mass
$m_{0}$ can be written as%
\begin{equation}
S_{m}=-\int m_{0}\sqrt{\tilde{g}_{\mu\nu}\tilde{u}^{\mu}\tilde{u}^{\nu}%
}d\tilde{s}+S_{int} \label{9}%
\end{equation}

where $S_{int}$ represents an action describing the interaction of the
particle with any nongravitational sources in its environment.

\bigskip

\ \ Dropping a total divergence in the action, the action can be rewritten as%
\begin{equation}
S=\int d^{4}x\sqrt{\tilde{g}}\ b\left\{  \frac{1}{2\kappa^{2}}\tilde{R}%
[\tilde{g}_{\mu\nu}]-\frac{\Lambda}{\kappa^{2}}\right\}  +S_{m}\label{11}%
\end{equation}

Notice that the action of (\ref{11}) contains a scalar field $b$ that is
nonminimally coupled to the Ricci scalar as appears in the term $b\tilde
{R}[\tilde{g}_{\mu\nu}]$, so that we have a 4D Jordan frame\ representation of
the theory. We can perform a conformal transformation of the metric in order
to recast the theory in the Einstein frame\ representation, where the scalar
field decouples from the curvature scalar, but then becomes coupled to the
matter sector\cite{Dicke62}.

\subsection{Conformal transformation to the Einstein frame}

\ \ Let us now define the Einstein frame metric $g_{\mu\nu}$ in terms of the
Jordan frame metric $\tilde{g}_{\mu\nu}$:%
\begin{equation}
g_{\mu\nu}=b\tilde{g}_{\mu\nu},\ \ \ g^{\mu\nu}=b^{-1}\tilde{g}^{\mu\nu
},\ \ \ \sqrt{g}=b^{2}\sqrt{\tilde{g}}\label{12}%
\end{equation}

The action in the Einstein frame representation now takes the form\cite{CGHW}%
,\cite{scatter}%
\begin{equation}
S=\int d^{4}x\sqrt{g}\left\{  \frac{1}{2\kappa^{2}}\left[  R[g_{\mu\nu}%
]+\frac{3}{2}b^{-2}g^{\mu\nu}(\nabla_{\mu}b)(\nabla_{\nu}b)\right]
-b^{-1}\frac{\Lambda}{\kappa^{2}}\right\}  +S_{m}[b^{-1}g_{\mu\nu},\cdot
\cdot\cdot]\label{13}%
\end{equation}

Now define the scalar field $\phi(x)$ by%
\begin{equation}
b=e^{a\phi},\ \ \ \ \ \phi=\frac{1}{a}\ln b,\ \ \ \ \ a=\sqrt{\frac{2}{3}%
}\ \kappa\label{14}%
\end{equation}

where $a$ is a constant that is determined by requiring the kinetic term in
the action to take the canonical form $\frac{1}{2}g^{\mu\nu}\nabla_{\mu}%
\phi\nabla_{\nu}\phi$. Then the scalar field $\phi$ is explicitly related to
the extra dimensional scale factor $b$ by%
\begin{equation}
\phi(x)=\sqrt{\frac{3}{2}}\frac{1}{\kappa}\ln b(x),\ \ \ \ \ b(x)=\exp\left[
\sqrt{\frac{2}{3}}\ \kappa\phi(x)\right]  \label{16}%
\end{equation}

and the action takes the form
\begin{subequations}
\label{17}%
\begin{align}
S &  =\int d^{4}x\sqrt{g}\left\{  \frac{1}{2\kappa^{2}}R[g_{\mu\nu}]+\frac
{1}{2}g^{\mu\nu}\nabla_{\mu}\phi\nabla_{\nu}\phi-e^{-a\phi}\frac{\Lambda
}{\kappa^{2}}\right\}  +S_{m}[e^{-a\phi}g_{\mu\nu},\cdot\cdot\cdot
]\label{17a}\\
&  =S_{R}+S_{\phi}+S_{m}\label{17b}%
\end{align}

with $e^{-a\phi}=b^{-1}$, and we define%
\end{subequations}
\begin{equation}
S_{R}=\int d^{4}x\sqrt{g}\frac{1}{2\kappa^{2}}R[g_{\mu\nu}],\ \ \ \ \ S_{\phi
}=\int d^{4}x\sqrt{g}\left[  \frac{1}{2}\partial^{\mu}\phi\partial_{\mu}%
\phi-V(\phi)\right]  ,\ \ \ \ V(\phi)=\frac{\Lambda}{\kappa^{2}}e^{-a\phi
}\label{18}%
\end{equation}

The scalar field $\phi$ has canonical mass dimension 1, and $\kappa\phi$ is dimensionless.

\bigskip

\ \ The Einstein frame action contains an Einstein-Hilbert term for gravity,
along with an action for the scalar field $\phi$. In addition, there is a
matter action where matter fields have an anomalous coupling to the scalar
field. In passing from the Jordan frame to the Einstein frame, the coupling of
the scalar to the gravitational sector has been shifted to a coupling of the
scalar to the matter sector.

\section{Nonrelativistic matter}

\ \ Before obtaining an appropriate Schr\"{o}dinger equation for a
nonrelativistic matter particle, we first focus on the nonrelativistic limit
of the action for a classical particle. The matter action given by (\ref{9})
can be rewritten in the Einstein frame. The free particle portion of the
action $S_{m}=S_{\text{free}}+S_{int}$ is given by%
\begin{equation}
S_{\text{free}}=-\int m_{0}d\tilde{s}=-\int m_{0}b^{-1/2}ds=-\int mds=-\int
m_{0}b^{-1/2}\sqrt{g_{\mu\nu}u^{\mu}u^{\nu}}ds\label{19}%
\end{equation}

This shows that the particle mass in the Einstein frame representation is
given by\cite{Dicke62}%
\begin{equation}
m(x)=m_{0}b^{-1/2}\label{20}%
\end{equation}

Furthermore, in the nonrelativistic limit $ds=dt$, $u^{i}\ll u^{0}=1$ and we
identify the nonrelativistic portion of the classical Lagrangian as%
\begin{equation}
L_{\text{free}}=-m_{0}b^{-1/2}(g_{\mu\nu}u^{\mu}u^{\nu})^{1/2}=-m_{0}%
b^{-1/2}(u_{0}u^{0}+g_{ij}u^{i}u^{j})^{1/2}=m_{0}b^{-1/2}\left[  \frac{1}%
{2}(\vec{u}\cdot\vec{u})-1\right]  \label{21}%
\end{equation}

where use has been made of $-g_{ij}u^{i}u^{j}=\vec{u}\cdot\vec{u}=|\vec
{u}|^{2}\ll1$.

\bigskip

\ Next, we write the nonrelativistic limit of the interaction term in $S_{m}$
as%
\begin{equation}
S_{int}=\int L_{int}d\tilde{s}=-\int U(\vec{x})d\tilde{s}=-\int U(\vec
{x})b^{-1/2}ds\approx-\int b^{-1/2}U(\vec{x})dt\label{22}%
\end{equation}

and $L_{int}=b^{-1/2}U(\vec{x})$, where $U(\vec{x})$ is an arbitrary potential
energy function defined in the Jordan frame, describing interactions of the
particle with nongravitational forces. The nonrelativistic limit of the matter
action is then%
\begin{equation}
S_{m}=\int Ldt=\int dtb^{-1/2}\left\{  \left[  \frac{1}{2}m_{0}\vec{u}%
\cdot\vec{u}\right]  -\left[  U(\vec{x})+m_{0}\right]  \right\}  \label{23}%
\end{equation}

The potential $U(\vec{x})$ can be redefined to absorb the constant $m_{0}$,
i.e., $U(\vec{x})+m_{0}\rightarrow U(\vec{x})$, and we can therefore write the
Lagrangian for the nonrelativistic particle in the Einstein frame
representation as%
\begin{equation}
L=b^{-1/2}\left[  \frac{1}{2}m_{0}\vec{u}\cdot\vec{u}-U(\vec{x})\right]
\equiv b^{-1/2}L_{0}\label{24}%
\end{equation}

where $L_{0}\equiv L|_{b=1}$. If we define $U_{0}(\vec{x})=2U(\vec{x})/m_{0}$,
then (\ref{24}) takes the form, in cartesian coordinates,%
\begin{equation}
L=\frac{1}{2}m(x)\left[  \delta_{ij}u^{i}u^{j}-U_{0}(\vec{x})\right]
\label{25}%
\end{equation}

which coincides with that in (\ref{3}), in the limit of a flat spacetime where
$g_{00}=1$, $g_{ij}=-\delta_{ij}$ (for cartesian coordinates).

\section{The flat space nonrelativistic theory}

\ \ We want to focus on a laboratory type of setting for a nonrelativistic
particle, i.e., a setting where the explicit curvature of spacetime can be
ignored. In other words, we consider a small enough region of space where it
becomes reasonable to approximate the spacetime as Minkowskian, with
$g_{\mu\nu}=\eta_{\mu\nu}$, where $\eta_{\mu\nu}$ is the flat space Minkowski
metric, appropriately expressed in terms of the coordinate system used. (For
cartesian coordinates we have $\eta_{00}=\eta^{00}=1$, $\eta_{ij}=\eta
^{ij}=-\delta_{ij}=-\delta^{ij}$.) In this approximation the curvature action
$S_{R}$ given by (\ref{18}) is set to zero, and the action $S$ of (\ref{17})
for the system reduces to
\begin{subequations}
\label{26}%
\begin{align}
S &  =S_{\phi}+S_{m}=\int d^{4}x\sqrt{\eta}\left[  \frac{1}{2}\eta^{\mu\nu
}\partial_{\mu}\phi\partial_{\nu}\phi-b^{-1}\frac{\Lambda}{\kappa^{2}}\right]
+S_{m}\label{26a}\\
&  =\int d^{4}x\sqrt{\eta}\left[  \frac{1}{2}\eta^{\mu\nu}\partial_{\mu}%
\phi\partial_{\nu}\phi-e^{-a\phi}\frac{\Lambda}{\kappa^{2}}\right]  +\int
dt\ e^{-a\phi/2}\left[  \frac{1}{2}m_{0}(-\eta_{ij})u^{i}u^{j}-U(\vec
{x})\right]  \label{26b}%
\end{align}

where (\ref{24}) has been used for the matter Lagrangian, and the metric
$\eta_{\mu\nu}$ has been left explicitly in the expressions, with a reminder
that the metric is one for flat spacetime, with an arbitrary choice for
cartesian or curvilinear coordinates. (We have $\eta_{00}=1$, and $\eta_{ij}$
is just the negative of the metric for a Euclidean space, with $-\eta_{ij}>0$
for $i\neq j$, and $\eta=|\det\eta_{\mu\nu}|$.) The parameterization
$b=e^{a\phi}$, as given by (\ref{14}), has also been used. A lagrangian
density $\mathcal{L}_{m}$ for the matter particle can be defined by writing
$\mathcal{L}_{m}=\sqrt{\eta}L\delta^{(3)}(\vec{x}-\vec{x}_{p})$ where $\vec
{x}_{p}$ locates the instantaneous position of the particle. Then (\ref{26})
can be expressed as%
\end{subequations}
\begin{align}
S &  =\int d^{4}x\sqrt{\eta}\left[  \frac{1}{2}\eta^{\mu\nu}\partial_{\mu}%
\phi\partial_{\nu}\phi-e^{-a\phi}\frac{\Lambda}{\kappa^{2}}\right]
\nonumber\\
&  +\int d^{4}x\sqrt{\eta}e^{-a\phi/2}\left[  \frac{1}{2}m_{0}(-\eta
_{ij})u^{i}u^{j}-U(\vec{x})\right]  \delta^{(3)}(\vec{x}-\vec{x}%
_{p})\label{27}%
\end{align}

\bigskip

\ \ From this action the equation of motion for the field $\phi$ can be
obtained:%
\begin{equation}
\square\phi+\frac{\partial V(\phi)}{\partial\phi}-\sigma=0;\ \ \ \ \ V(\phi
)=e^{-a\phi}\frac{\Lambda}{\kappa^{2}},\ \ \ \ \ \sigma=\frac{\partial
\mathcal{L}_{m}}{\partial\phi}=\left(  -\frac{a}{2}\right)  \sqrt{\eta}%
L\delta^{(3)}(\vec{x}-\vec{x}_{p})\label{28}%
\end{equation}

where $\square=\nabla_{\mu}\partial^{\mu}=\partial_{t}^{2}-\nabla^{2}$. (The
field $\phi$ is referred to as a 4D \textquotedblleft
dilaton\textquotedblright\ field.) Since $\mathcal{L}_{m}\propto\delta
^{(3)}(\vec{x}-\vec{x}_{p})$, we take the $\sigma$ term to vanish at all
$\vec{x}\neq\vec{x}_{p}$. So, at least as a first approximation, we set
$\sigma=0$ in the equation of motion in order to obtain a background solution
for $\phi(x)$ that can couple to the matter particle in $L$. The equation of
motion for $\phi$ therefore reads%
\begin{equation}
\square\phi-a\frac{\Lambda}{\kappa^{2}}e^{-a\phi}=0\label{29}%
\end{equation}

\bigskip

\bigskip

\ \ Here we focus attention on static solutions of (\ref{29}), that satisfy%
\begin{equation}
\nabla^{2}\phi=-a\frac{\Lambda}{\kappa^{2}}e^{-a\phi}\label{30}%
\end{equation}

Solutions to this equation can be found for certain specific cases.

\bigskip

As an example, we consider $\phi(x,y)$ to be a function of the two cartesian
coordinates $x$ and $y$, and (\ref{30}) reads as%
\begin{equation}
(\partial_{x}^{2}+\partial_{y}^{2})\phi=-a\frac{\Lambda}{\kappa^{2}}e^{-a\phi
}=2\tilde{\kappa}\frac{\Lambda}{\kappa^{2}}e^{2\tilde{\kappa}\phi}\label{31}%
\end{equation}

where $\tilde{\kappa}\equiv-\frac{1}{2}a$. Then (\ref{31}) is recognized as
the 2D Liouville differential equation\cite{Liouville},\cite{Crowdy} (also
see, for example,\cite{Gibbons} and \cite{DHoker}). Upon defining $\zeta
=x+iy$, the solutions are given by\cite{Gibbons},\cite{Crowdy}%
\begin{subequations}
\begin{align}
b^{-1} &  =e^{2\tilde{\kappa}\phi}=\frac{2\kappa^{2}}{\tilde{\kappa}%
^{2}|\Lambda|}\frac{|f^{\prime}(\zeta)|^{2}}{\left[  |f(\zeta)|^{2}+1\right]
^{2}},\ \ \ (\Lambda=-|\Lambda|<0)\label{32a}\\
b^{-1} &  =e^{2\tilde{\kappa}\phi}=\frac{2\kappa^{2}}{\tilde{\kappa}%
^{2}|\Lambda|}\frac{|f^{\prime}(\zeta)|^{2}}{\left[  |f(\zeta)|^{2}-1\right]
^{2}},\ \ \ (\Lambda=+|\Lambda|>0)\label{32b}%
\end{align}

where $f(\zeta)$ is a holomorphic function of $\zeta$, and $f^{\prime}%
(\zeta)=df/d\zeta$. The matter coupling function $b^{-\frac{1}{2}}%
=e^{\tilde{\kappa}\phi}$ that appears in the matter particle lagrangian $L$ is%
\end{subequations}
\begin{equation}
b^{-\frac{1}{2}}=e^{\tilde{\kappa}\phi}=\frac{\sqrt{2}\kappa}{|\tilde{\kappa
}|\sqrt{|\Lambda|}}\frac{|f^{\prime}(\zeta)|}{\left[  |f(\zeta)|^{2}%
\pm1\right]  },\ \ (\Lambda=\mp|\Lambda|)\label{33}%
\end{equation}

Different choices for the function $f(\zeta)$ give different mathematical
solutions to Liouville's equation. (See, e.g.,\cite{Gibbons},\cite{MorrisPLB}
for some applications to dilaton gravity and low energy string theory.)

\section{Quantization}

\ \ The classical lagrangian for a particle in the flat space nonrelativistic
limit is given by (\ref{24}), for instance. In terms of cartesian coordinates,
we can write the lagrangian as%
\begin{equation}
L=b^{-1/2}\left[  \frac{1}{2}m_{0}\delta_{ij}u^{i}u^{j}-U(\vec{x})\right]
=T-\mathcal{U};\ \ \ \ \ \mathcal{U}=b^{-1/2}U(\vec{x})\label{34}%
\end{equation}

(For cartesian coordinates the spatial metric, with positive signature is
$-g_{ij}=\delta_{ij}$, but one could use other coordinates with an appropriate
metric $\delta_{ij}\rightarrow\gamma_{ij}$.)

As noted in (\ref{20}), the effective mass of the particle in the Einstein
frame representation is%
\begin{equation}
m(\vec{x})=m_{0}b^{-1/2}=m_{0}e^{-a\phi/2}\label{35}%
\end{equation}

where we assume that $b=b(\vec{x})$ is time independent. The classical kinetic
energy is%
\begin{equation}
T=\frac{1}{2}m_{0}b^{-1/2}\delta_{ij}u^{i}u^{j}=\frac{1}{2}m(\vec{x}%
)\delta_{ij}u^{i}u^{j}=\frac{1}{2}m(\vec{x})\vec{u}\cdot\vec{u}\label{36}%
\end{equation}

The canonically conjugate momentum is
\begin{subequations}
\label{37}%
\begin{align}
p^{i} &  =p_{i}=\frac{\partial L}{\partial u^{i}}=m_{0}b^{-1/2}u^{i}=m(\vec
{x})u^{i};\label{37a}\\
u^{i} &  =\frac{p^{i}}{m(\vec{x})}=b^{1/2}\frac{p^{i}}{m_{0}}\label{37b}%
\end{align}

Therefore, in terms of the canonical momentum, the kinetic energy is%
\end{subequations}
\begin{equation}
T=b^{1/2}\delta_{ij}\frac{p^{i}p^{j}}{2m_{0}}=b^{1/2}\frac{\vec{p}\cdot\vec
{p}}{2m_{0}}\label{38}%
\end{equation}

\bigskip

\ \ We can now implement the quantization procedure described in the
Introduction, and used in previous works\cite{cari},\cite{refe},\cite{axel1}%
,\cite{epjp}. We write $T$ in a symmetrical form%
\begin{equation}
T=\frac{1}{2m_{0}}(b^{1/4}\vec{p})\cdot(b^{1/4}\vec{p})\label{39}%
\end{equation}

and then make the canonical quantum replacement $\vec{p}\rightarrow
-i\hbar\nabla$ in order to obtain the kinetic part of the quantum Hamiltonian
$\hat{H}$:%
\begin{equation}
T\rightarrow\hat{T}=-\frac{\hbar^{2}}{2m_{0}}(b^{1/4}\nabla)\cdot
(b^{1/4}\nabla)=-\frac{\hbar^{2}}{2m_{0}}\left[  b^{1/2}\nabla^{2}%
+b^{1/4}(\nabla b^{1/4})\cdot\nabla\right]  \label{40}%
\end{equation}

The classical Hamiltonian%
\begin{equation}
H=T+b^{-1/2}U(\vec{x})=T+\mathcal{U}(\vec{x})\label{41}%
\end{equation}

is then replaced by a quantum Hamiltonian%
\begin{equation}
\hat{H}=\hat{T}+\mathcal{U}(\vec{x})=\hat{T}+b^{-1/2}U(\vec{x})\label{42}%
\end{equation}

where $\hat{T}$ is given by (\ref{40}).

\section{Application: A dilaton string background}

\subsection{The nonrelativistic quantum particle}

\ \ As an example we consider a particle subjected to harmonic oscillation in
the $x$-$y$ plane with translation invariance in the $z$ direction, or a
particle subject to oscillation in a two dimensional space of the $x$-$y$
plane. We employ the quantization procedure described in the previous section.
Consider now a solution of the 2D Liouville equation for which $f(\zeta
)=A\zeta$, where $\zeta=x+iy$, and hence $f^{\prime}(\zeta)=A$, $|f(\zeta
)|^{2}=A^{2}(x^{2}+y^{2})=A^{2}r^{2}$. Then (for $\Lambda=-|\Lambda|$)
(\ref{33}) gives a solution (see, for example, \cite{Gibbons} and
\cite{MorrisPLB})%
\begin{equation}
b^{-1/2}=\frac{C}{(1+A^{2}r^{2})},\ \ \ \ C=\frac{\sqrt{2}\kappa A}%
{|\tilde{\kappa}|\sqrt{|\Lambda|}};\ \ \ \ \ b^{1/2}=\frac{(1+A^{2}r^{2})}%
{C};\ \ \ \ \ b^{1/4}=\frac{(1+A^{2}r^{2})^{1/2}}{\sqrt{C}}\label{43}%
\end{equation}

Using%
\begin{equation}
(\partial_{r}b^{1/4})=\frac{(1+A^{2}r^{2})^{-1/2}}{\sqrt{C}}\cdot
A^{2}r\label{44}%
\end{equation}

(\ref{40}) yields%
\begin{equation}
\hat{T}=-\frac{\hbar^{2}}{2m_{0}}\left[  \frac{(1+A^{2}r^{2})}{C}\nabla
^{2}+\frac{A^{2}r}{C}\partial_{r}\right]  \label{45}%
\end{equation}

Interestingly, this is the same form of the kinetic Hamiltonian that has been
studied by various authors for a nonlinear quantum oscillator with position
dependent mass\cite{cari},\cite{refe},\cite{axel1},\cite{epjp}. In fact, if we
choose a potential $U(r)=\frac{1}{2}Kr^{2}$, we have a potential term in the
Hamiltonian $\hat{H}$ given by
\begin{equation}
\mathcal{U}(r)=b^{-1/2}U(r)=\frac{CK}{2}\frac{r^{2}}{(1+A^{2}r^{2})}\label{47}%
\end{equation}

Upon choosing $C=1$, $A^{2}=\lambda>0$, and $K=m_{0}\alpha^{2}$, the
Hamiltonian is \cite{note}%
\begin{equation}
\hat{H}=-\frac{\hbar^{2}}{2m_{0}}\left[  (1+\lambda r^{2})\nabla^{2}+\lambda
r\partial_{r}\right]  +\frac{1}{2}\frac{m_{0}\alpha^{2}r^{2}}{(1+\lambda
r^{2})} \label{48}%
\end{equation}

This system and related systems have been studied extensively for the cases of
1, 2, 3, and $n$ dimensions of space\cite{cari},\cite{refa},\cite{refe}%
,\cite{axel1},\cite{jmp},\cite{epjp}. See, e.g., \cite{cari} and
\cite{axel1}-\cite{epjp} for exact expressions for the normalizable,
orthogonal eigenfunctions, along with the spectrum of energy eigenvalues for a
class of Hamiltonians of this type. In those works, however, the mass function
$m(\vec{x})$ is an assumed input function, whereas here it emerges as a
solution of the 2D Liouville equation associated with the dilaton field
$\phi(r)$.

\bigskip

\ \ Previous studies of this system have focused largely on its mathematical
properties, with some mention that such systems may achieve physical
realizations in condensed matter settings, as pointed out in\cite{cari}, for
instance. Here, however, we see this type of system being realized in the
context of an inhomogeneous compactification of extra space dimensions,
resulting in a quantum particle in \textquotedblleft
ordinary\textquotedblright\ space having an effective position dependent mass,
whose variation depends upon a variation of the dilaton field, or
equivalently, a variation in the size of the extra dimensions. The effective
mass is%
\begin{equation}
m(r)=m_{0}b^{-1/2}(r)=\frac{m_{0}C}{(1+A^{2}r^{2})}\label{49}%
\end{equation}

in an effective potential%
\begin{equation}
\mathcal{U}(r)=b^{-1/2}(r)U(r)\propto\frac{r^{2}}{(1+A^{2}r^{2})}\label{50}%
\end{equation}

There is a discrete spectrum of bound states\cite{epjp}, where we could
consider nonrelativistic particles as becoming trapped within the
\textquotedblleft dilatonic\textquotedblright\ string-like core described by
the dilaton coupling function $b^{-1/2}$, for which the size of the extra
dimensions is%
\begin{equation}
b(r)\sim(1+A^{2}r^{2})^{2}\label{51}%
\end{equation}

So, the coupling function $b^{-1/2}=e^{\tilde{\kappa}\phi}$ has a maximum in
the string core where $b\sim1$, and outside the core the coupling function
$b^{-1/2}(r)$ decreases, and the extra dimensional scale factor $b(r)$
increases (i.e., the extra dimensions get bigger). Bound state particles are
then localized within the dilatonic string core, where the extra dimensions
are small, and the effective particle mass is large.

\subsection{The dilaton string}

\ \ We choose a negative cosmological constant, $\Lambda=-|\Lambda|$, so that
from (\ref{28}) the dilaton string has a potential given by%
\begin{equation}
V(\phi)=-\frac{|\Lambda|}{\kappa^{2}}e^{-a\phi}=-\frac{|\Lambda|}{\kappa^{2}%
}e^{2\tilde{\kappa}\phi}=-\frac{|\Lambda|}{\kappa^{2}}b^{-1}\label{52}%
\end{equation}

The energy density of the scalar field $\phi$ is%
\begin{equation}
\mathcal{H}=\mathcal{H}_{\text{kin}}+\mathcal{H}_{\text{pot}}=\frac{1}%
{2}(\partial_{r}\phi)^{2}+V(\phi) \label{53}%
\end{equation}

with $r=\sqrt{x^{2}+y^{2}}$ being the radial distance from the center of the
vortex (2D space dimensions) or the center of the string (which is centered on
the $z$-axis in a 3D space). We now adopt the settings $C=1$ and
$A^{2}=\lambda>0$. We then have%
\begin{equation}
b=e^{a\phi}=(\lambda r^{2}+1)^{2},\ \ \phi=-\frac{1}{\tilde{\kappa}}%
\ln(\lambda r^{2}+1)=-\frac{1}{\tilde{\kappa}}\ln(\xi+1),\ \ \ \ \xi
\equiv\lambda r^{2}\label{54}%
\end{equation}

where $\xi=\lambda r^{2}$ is a dimensionless distance parameter. The energy
per unit length of the string i.e., the string tension, is (assuming, for
simplicity, $z$ independence)
\begin{equation}
\mu=\mu_{\text{kin}}+\mu_{\text{pot}}=2\pi\int_{0}^{r_{C}}(\mathcal{H}%
_{\text{kin}}+\mathcal{H}_{\text{pot}})\ rdr=2\pi\int_{0}^{r_{C}}%
\mathcal{H}(r)\ rdr=\frac{\pi}{\lambda}\int_{0}^{\xi_{C}}\mathcal{H}(\xi
)d\xi\label{55}%
\end{equation}

where $r_{C}$ and $\xi_{C}$ are large distance cutoffs, as the tension
diverges logarithmically, as with a global cosmic string\cite{Vilenkin}%
,\cite{VSbook}. (This dilaton string resembles the string studied
in\ \cite{Gibbons} and \cite{MorrisPLB}, with the cosmological constant
replacing a constant magnetic $H$ field.) Using (\ref{52})-(\ref{55}), some
calculation yields%
\begin{equation}
\frac{\mu}{2\pi}=\frac{1}{\tilde{\kappa}^{2}}\left[  \ln(\xi_{C}+1)+\frac
{1}{(\xi_{C}+1)}-1\right]  -\frac{|\Lambda|}{2\lambda\kappa^{2}}\left[
1-\frac{1}{(\xi_{C}+1)}\right]  \label{56}%
\end{equation}

For $\xi_{C}\gg1$ ($r_{C}\gg1/\sqrt{\lambda}$) this simplifies to%
\begin{equation}
\frac{\mu}{2\pi}\approx\frac{1}{\tilde{\kappa}^{2}}\left(  \ln\xi
_{C}-1\right)  -\frac{|\Lambda|}{2\lambda\kappa^{2}} \label{57}%
\end{equation}

\bigskip

\ \ We note that since the potential is negative, but the kinetic term is
positive, one can find%
\begin{equation}
\Big|\frac{\mathcal{H}_{\text{kin}}}{\mathcal{H}_{\text{pot}}}\Big|=\frac
{\frac{1}{2}(\partial_{r}\phi)^{2}}{|V(\phi)|}=\frac{2\lambda^{2}\kappa^{2}%
}{|\Lambda|\tilde{\kappa}^{2}}r^{2} \label{58}%
\end{equation}

so that the energy density becomes negative within a certain critical radius:%
\begin{equation}
\mathcal{H}\leq0\ \ \ \ \text{for\ \ \ \ }r\leq r_{crit}=\frac{|\Lambda
|^{1/2}|\tilde{\kappa}|}{\sqrt{2}\lambda\kappa},\ \ \ \ \ \ \xi_{crit}%
=\frac{|\Lambda|\tilde{\kappa}^{2}}{2\lambda\kappa^{2}} \label{59}%
\end{equation}

so that inside this portion of the string core the weak energy condition is
violated. However, the string tension is positive, $\mu>0$, provided that
$\xi_{C}\gg1$, and the condition%
\begin{equation}
\ln(\xi_{C}+1)>1+\xi_{crit} \label{60}%
\end{equation}

is satisfied.

\section{Summary}

\ \ Beginning with a classical particle propagating in a four dimensional
subspace of a five dimensional spacetime, where the internal space dimension
gets inhomogeneously, toroidally, compactified, we arrive at an effective 4
dimensional theory for a classical particle. When a conformal transformation
is made from the resulting 4D Jordan frame to a 4D Einstein frame, the
particle acquires a position dependent mass. The classical nonrelativistic
system in a flat Einstein frame spacetime can be quantized using existing
quantization procedures. The result is the emergence of a quite general class
of quantum systems that have position dependent mass and somewhat arbitrary
potentials. Systems of this type, e.g., the nonlinear oscillator with a PDM,
emerging from different physical circumstances, have been studied already.
Here, we show another way in which this quantum system and generalizations of
it can arise. It is also seen that within the type of scenario considered here
there is a somewhat different physical interpretation involving a particle
interacting with a dilaton field, which is associated with the size of the
extra dimension.

\bigskip

\ \ The hope is that a possible physical realization of such systems, in the
context of higher dimensional physics, will provide motivation for their
further study, along with new physical interpretations of results
materializing from these systems. We have presented an application of this
formalism to obtain a previously studied system describing a nonlinear quantum
oscillator with a position dependent mass. In the scenario presented here,
however, the effective position dependent particle mass depends explicitly
upon the scalar dilaton field (i.e., the size of the extra dimension), which
itself may have nontrivial, interesting structure.


\begin{thebibliography}{99}
\bibitem {note}Our choice $C=1$ fixes the ratio $\lambda/|\Lambda|$, but we do
not concern ourselves with this. This restriction could be obviated with a
redefined constant $K=Cm_{0}\alpha^{2}$ along with a redefinition of the mass
parameter, $Cm_{0}\rightarrow m_{0}$.
\end{thebibliography}
\end{document}